\documentclass[prl,twocolumn,epsfig,superscriptaddress,rotate,showpacs]{revtex4}
\usepackage{epsfig}
\usepackage{graphicx}
\usepackage{amsmath}

\newcommand{\beq}{\begin{eqnarray}}
\newcommand{\eeq}{\end{eqnarray}}
\begin{document}
\title{Many-Body Coherent Destruction of Tunneling} 
\author{Jiangbin Gong}\email{phygj@nus.edu.sg}
\affiliation{Department of Physics and Center for Computational
Science and Engineering, National University of Singapore, 117542,
Singapore} \affiliation{NUS Graduate School for Integrative Sciences
and Engineering, Singapore 117597, Singapore}
\author{Luis Morales-Molina}
\affiliation{Department of Physics and Center for Computational
Science and Engineering, National University of Singapore, 117542,
Singapore}
\author{Peter H\"{a}nggi}
\affiliation{Department of Physics and Center for Computational
Science and Engineering, National University of Singapore, 117542,
Singapore} \affiliation{Theoretische Physik I, Institut f\"{u}r
Physik, Universit\"{a}t Augsburg, D - 86135 Augsburg, Germany}

\begin{abstract}
A new route to coherent destruction of tunneling is established by
considering a monochromatic fast modulation of the self-interaction
strength of a many-boson system. The modulation can be tuned such
that only an arbitrarily, {\em a priori}  prescribed number of
particles are allowed to tunnel. The associated tunneling dynamics
is sensitive to the odd or even nature of the number of bosons.
\end{abstract}
\date{\today}
\maketitle

The phenomenon of coherent destruction of tunneling (CDT)
\cite{CDT1991,Hanggirev} by a driving field has been one seminal
result in studies of quantum dynamics control. Direct observation of
CDT was recently achieved in several experiments
\cite{devalle,oberthaler}, one of which \cite{oberthaler} involving
non-interacting cold atoms in a double-well potential.  CDT in
interacting many-body systems has also attracted considerable
interest, with previous work focusing on driving fields on resonance
with the interaction energy \cite{creffield,holthaus}. The
Mott-superfluid transition in ultracold systems via a mechanism
similar to single-particle CDT has also been observed experimentally
\cite{Erimondo}.

Traditionally, the driving field in CDT studies is to directly
modulate the bare single-particle levels of an undriven system. For
example, in a two-level theory of CDT, the driving field is to
modulate the energy difference between two bare levels. By contrast,
in this work we expose a new route to CDT by taking advantage of the
particle-particle interaction  in a many-body system. Specifically,
we consider the CDT in a two-mode Bose-Hubbard model that describes
a two-mode Bose-Einstein condensate (BEC).  We show, both
analytically and computationally, that a monochromatic off-resonance
driving of the self-interaction strength of the BEC can induce
different types of CDT, without a direct modulation of the
mode-energy bias. Interestingly, this makes it possible to precisely
control, at least in principle, the number of bosons allowed to
tunnel.
Another remarkable prediction is the sensitivity of the full-quantum
dynamics to the even or odd nature of the number of bosons. Note
that in other contexts such as matter-wave solitons, intriguing
effects of a periodic modulation of the self-interaction strength of
a BEC (the so-called ``Feshbach-resonance management") have been
discovered \cite{kevre}, but on the mean-field level only.

Consider then the following Bose-Hubbard Hamiltonian for a two-mode
BEC,
\begin{eqnarray}
H=v\hbar (a_{l}^{\dagger}a_r+ a_{r}^{\dagger}a_l)/2+
g(t)\hbar(a_{l}^{\dagger}a_l-a_{r}^{\dagger}a_r)^2/4,
\label{Hami}
\end{eqnarray}
where $r$ and $l$ are mode indices, $a_{k}$ and $a_{k}^{\dagger}$
($k=r,l$) are the bosonic annihilation and creation operators, $v$
describes the constant tunneling rate between the two modes, and
$g(t)$ is the interaction strength between same-mode bosons.  We use
the unit of $v$ to appropriately scale all the parameters such that
$v$, $g(t)$, and $t$ all become dimensionless variables.  The total
number of bosons $N=a_{l}^{\dagger}a_l+a_r^{\dagger}a_r$ is a
conserved quantity and the dimension of the Hilbert space is $N+1$.
Using the Schwinger representation of angular
 momentum operators, namely, $J_{x}=
(a_{l}^{\dagger}a_r+a_{r}^{\dagger}a_l)/2$,
 $J_y=(a_{r}^{\dagger}a_l-a_{l}^{\dagger}a_r)/(2i)$, and
 $J_z=(a_{l}^{\dagger}a_l-a_{r}^{\dagger}a_r)/2$, eq. (\ref{Hami})
 reduces to
 \begin{eqnarray}
H(t)= v\hbar J_x+ g(t)\hbar J_z^2. \label{Hami2}
\end{eqnarray}
The Hilbert space is expanded by the eigenstates of $J_z$,  denoted
$|m\rangle$, with $J_z|m\rangle=m|m\rangle$. The mode population
difference is given by the expectation value
of $2J_{z}$. For later use we also define $J_{+}\equiv J_{x}+iJ_y$.
Note that there is no energy bias between the two modes, and that
the time dependence of the Hamiltonian arises from $g(t)$, which is
assumed to be \cite{kevre}
\begin{eqnarray}
g(t)=g_0+g_1\cos(\omega t). \label{gt}
\end{eqnarray}
Under appropriate conditions $H(t)$ describes a BEC
distributed in the two wells of a double-well potential
\cite{oberthaler}, in the ground band and the first-excited band
associated with an accelerating optical lattice \cite{qi}, or in two
hyperfine levels. For convenience, below we focus on the
first context, which can be best realized by optical superlattices
\cite{naturepapers}.
 Hence $r$($l$) denotes the right (left) well. Our
central idea is to use an off-resonance oscillation in
$g(t)$ to switch off the tunneling between the left and right wells.

The Floquet
operator associated with $H(t)$ is given by $\hat{F}\equiv {\cal T} \left[
\exp(-i\int_{0}^{2\pi/\omega}H(t')/\hbar\ dt')\right]$, where ${\cal
T}$ is the time-ordering operator. Its eigenstates are the Floquet states,
with eigenvalues $\exp(-i\epsilon 2\pi
/\hbar\omega)$, where $\epsilon$ is the quasi-energy. Because $H(t)$
apparently possesses a parity symmetry, namely, it is invariant upon
an exchange of the indices $l$ and $r$, the Floquet states can be
chosen as either positive-parity or negative-parity states. If two
opposite-parity Floquet states cross or touch, then similar to the
single-particle CDT mechanism, their superposition, which is still a
Floquet state, breaks the left-right symmetry and hence CDT occurs
\cite{korsch}.

Figure 1 {depicts} that the degeneracy between opposite-parity
states can easily occur, thus suggesting that CDT is possible via
solely a monochromatic modulation in the self-interaction strength.
Next we examine some detailed features presented in Fig. 1, for
$g_0=0$, $N=10$, and $v=1$ as an example. In Fig. 1(a),
$\omega=40\gg v=1$, and three values of $g_1/\omega$ for which level
degeneracies occur are explicitly marked by vertical dashed lines.
At point I, the level crossing involves two positive-parity states
and one negative-parity state. At point II, two pairs of
opposite-parity states become degenerate simultaneously. At point
III, three states in the middle of the Floquet spectrum cross, and
two additional pairs of opposite-parity Floquet states also touch
each other. Similar spectral patterns can be found in other regimes
of $g_{1}/\omega$. In Fig. 1(b), $\omega=2\pi$ is quite comparable
to $v=1$. In this intermediate-frequency case the spectral patterns
become more complicated. Nevertheless, as indicated by those
vertical dashed lines in Fig. 1(b), when $g_1$ increases, the
spectral pattern in Fig. 1(b) becomes analogous to those seen in
Fig. 1(a).


\begin{figure}
\begin{center}
\begin{tabular}{lc}
\includegraphics[width=6.0cm,height=4.5cm]{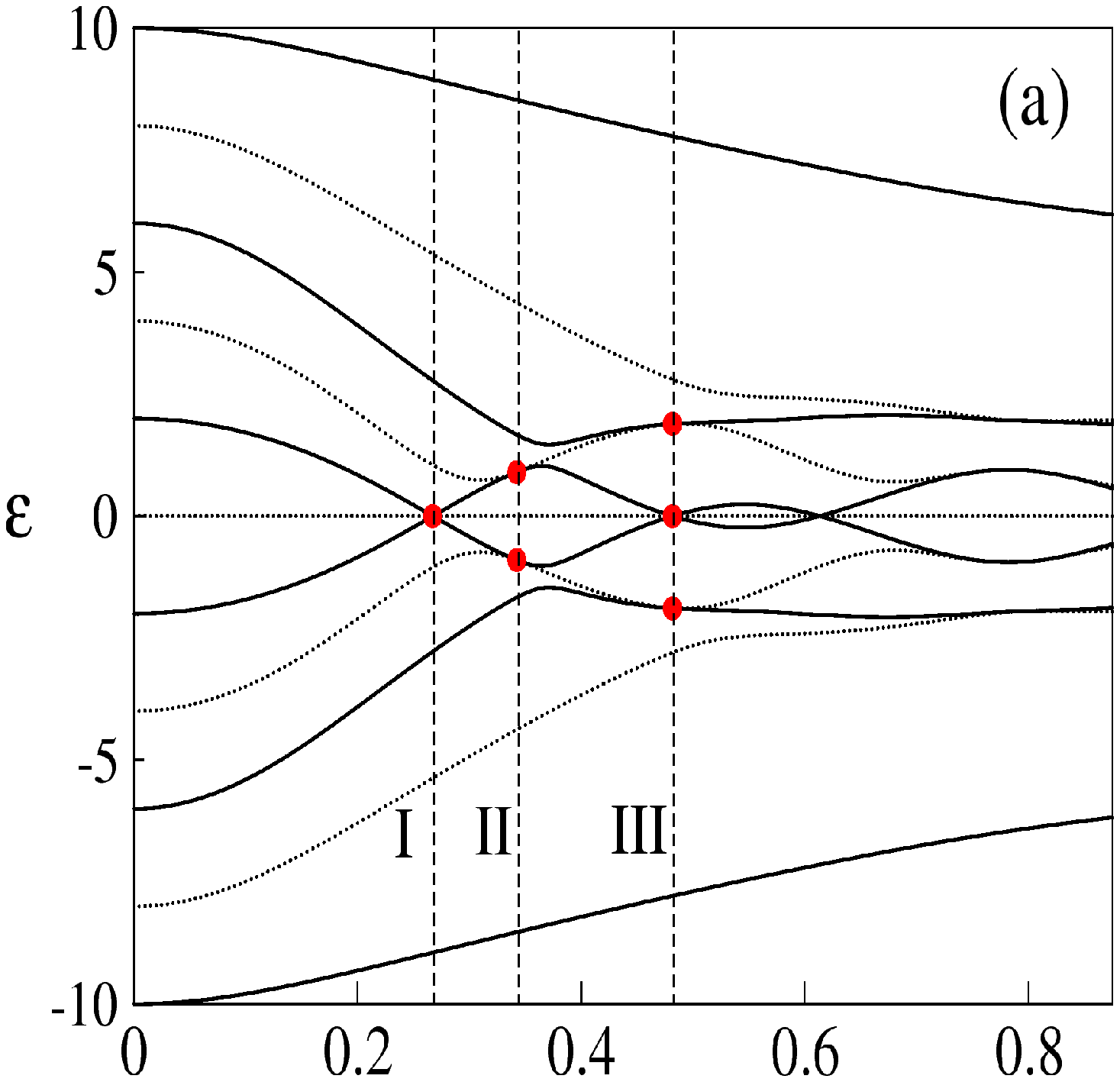}
\\

\includegraphics[width=6.0cm,height=4.5cm]{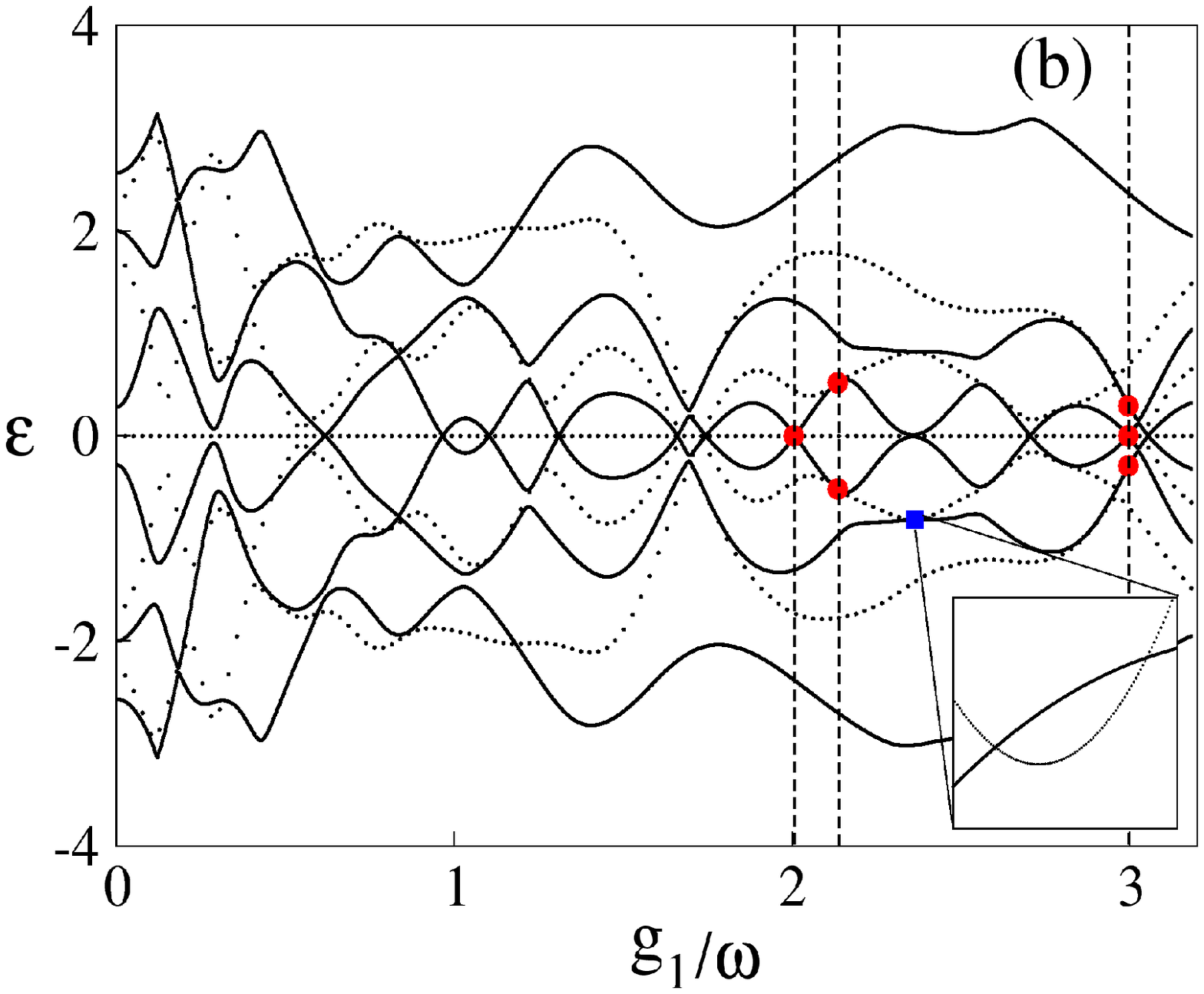}
\end{tabular}
\caption{(Color online) Quasienergy spectrum versus
$g_{1}/\omega$. Vertical dashed lines indicate the location {(i.e. $
{g_1/\omega \sim 0.267,0.344, 0.481}$)} of level degeneracies.
Dotted (solid) lines are for states
  with negative (positive) parity.
$N=10$, $g_0=0$, and $v=1$. In (a)
  $\omega=40$ and
in (b)
  $\omega=2\pi$.  Spectral details around the square are shown in inset in (b), with two level-crossings. Here and in other figures all variables are dimensionless.} \label{Fig:spectrum}
\end{center}
\end{figure}


We develop below a theory for the Floquet spectrum in the
high-frequency regime, with its validity condition elaborated later.
In that regime, the full Floquet theory can be expanded to the first
order of $1/\omega$ \cite{Hanggirev} and a static effective
Hamiltonian $H_{\text{eff}}$ for the driven quantum dynamics can be
obtained by averaging out the driving-field effects \cite{kohler}.
Explicitly,
\begin{eqnarray}
H_{\text{eff}}=\frac{\omega}{2\pi}\int^{2\pi/\omega}_{0}e^{[iA(t)J_z^2]}
(g_0\hbar J_z^2+v\hbar J_x) e^{[-iA(t)J_z^2]}\ dt, \label{Heff}
\end{eqnarray}
where $A(t)=\int^{t}_{0}g_{1} \cos(\omega t) \
dt=(g_1/\omega)\sin(\omega t)$.  Using the identity in the SU(2)
algebra, i.e., $
 e^{iA(t)J_z^2} J_x e^{-iA(t)J_z^2}  =
(J_{+}/2)e^{iA(t)(2J_z+1)}  + \text{c.c.}, \label{su2}
$
  where c.c. means complex conjugate of the preceding term, and
substituting this into Eq. (\ref{Heff}) to perform the integral, we
obtain the effective Hamiltonian,
\begin{eqnarray}
H_{\text{eff}}=g_0\hbar J_z^2+ (v\hbar J_{+}/2) {\cal J}
_{0}\left[g_1(2J_z+1)/\omega\right] + \text{c.c.}, \label{heff1}
\end{eqnarray} where ${\cal J}_0(x)$ is the ordinary Bessel function
of order zero. Equation (\ref{heff1}) indicates that the net effect
of a fast modulation in $g(t)$ is the rescaling factor ${\cal
J}_{0}[g_1(2J_z+1)/\omega]$, which depends on $J_z$, i.e., the
population difference between the two wells. Further, a nonzero
$g_0$ leads to the $g_0\hbar J_z^2$ term in $H_{\text{eff}}$. It is
well known that this term can induce population localization via a
self-trapping mechanism. To isolate population localization due to
possible CDT phenomena from that due to self-trapping, we will not
consider cases with nonzero $g_0$ until much later.

In the eigen-representation of $J_z$, $H_{\text{eff}}$ is a
tri-diagonal matrix.  Further, if $\langle
m-1|H_{\text{eff}}|m\rangle=0$, then $\langle
m|H_{\text{eff}}|m-1\rangle=0$ and we must also have $\langle
1-m|H_{\text{eff}}|-m\rangle=\langle -m |H_{\text{eff}}|1-m\rangle
=0$ due to symmetry consideration. These four zero matrix elements
divide the tri-diagonal matrix of $H_{\text{eff}}$ into three
uncoupled subspaces, i.e.,
\begin{eqnarray}
H_{\text{eff}} = \left[\begin{array}{ccc} h_{l} & 0 & 0 \\ 0 & h_{i}
&
0\\ 0& 0& h_{r}\\
\end{array}\right],
\end{eqnarray}
where $h_{l}$ represents a sub-matrix of $H_{\text{eff}}$ in the
subspace spanned by states $|N/2\rangle$, $|N/2-1\rangle$, $\cdots$,
$|m\rangle$ (assuming $m>0$) , $h_{r}$ represents a sub-matrix of
$H_{\text{eff}}$ in the subspace spanned by states $|-N/2\rangle$,
$|-N/2+1\rangle$, $\cdots$, $|-m\rangle$, and $h_{i}$ represents the
third block matrix involving other remaining basis states.
Therefore, if $\langle m-1|H_{\text{eff}}|m\rangle =0$, then (i) the
transition between $|m\rangle$ ($|-m\rangle$) and all other basis
states $|m'\rangle $ ($|-m'\rangle$) with $m'<m$ will not occur;
(ii) the Floquet states must display degeneracy because $h_{r}$ is
identical with $h_{l}$ due to the parity symmetry of $H(t)$; and
(iii) the dimension of $h_{r}$ or $h_{l}$, namely, $(N/2-m+1)$, also
gives the expected number of degenerate pairs.

Suppose there are $N-i$ ($i$) particles in the left (right) well.
Without loss of generality we assume $i<N/2$. The associated quantum
state is given by $|m\rangle=|N/2-i\rangle$. Using Eq.
(\ref{heff1}), one finds that the condition $\langle
m-1|H_{\text{eff}}|m\rangle=0$ is equivalent to \begin{equation}
{\cal J} _{0}\left[{g_1(N-(2i+1))/\omega}\right]=0. \label{con2}
\end{equation}
As such, if $g_1(N-(2i+1))/\omega $ is tuned to become one root of
${\cal J}_0(x)$,  then the tunneling of one more particle from the
left well to the right well (hence $m\rightarrow m-1$) becomes
{prohibited} if the left well has already released $i$ particles to
the right well.

\begin{figure}
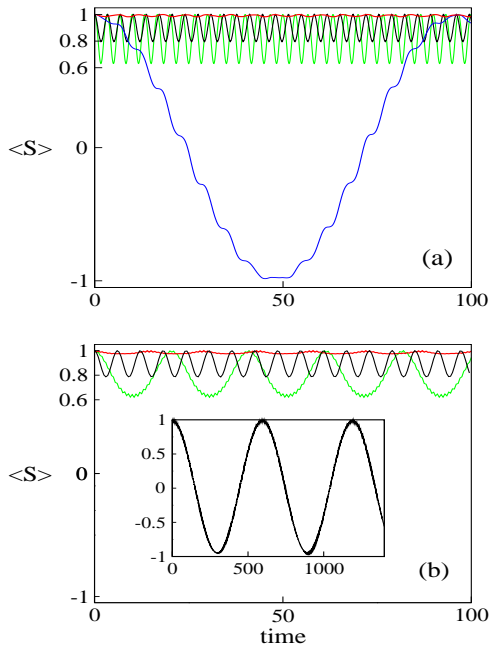

\begin{center}
\begin{tabular}{lc}
\includegraphics[width=6.4cm,height=4.cm]{fig2a.eps}
\\
\\
\includegraphics[width=6.4cm,height=4.cm]{population-imbalance-omega-2pi.eps}
\end{tabular}
\caption{(Color online) (a) Time dependence of $\langle S \rangle$
for $g_0=0$ and $N=10$, with $\omega=40$ in (a) and $\omega=2\pi$ in
(b). From top of bottom, values of $g_{1}/\omega$ are associated
with the three vertical lines (from left to right) in Fig.
\ref{Fig:spectrum}(a) for panel (a) and in Fig.
\ref{Fig:spectrum}(b) for panel (b).  For comparison, a case with
$g_1/\omega=10.5/40 {=0.2625}$ is also shown in (a). Inset of (b) is
for the first level crossing shown in the inset of Fig.
\ref{Fig:spectrum}(b). 
} \label{Fig:population}
\end{center}
\end{figure}

We now compare the theoretical result of Eq. (\ref{con2}) with our
computationally {precise} results in Fig. 1. For the cases in Fig.
1(a), theoretically the spectral degeneracy is expected to occur
when $g_1[10-(2i+1)]/\omega$ becomes a root of ${\cal J}_{0}(x)$. In
particular,  for the first root of ${\cal J}_0(x)$ at $x\sim 2.405$
and for $i=0,1,2$, the predicted degeneracy is at $g_{1}/\omega
\approx 0.267, 0.344$ and 0.481, with the pair number of level
degeneracies in each case given by the dimension of $h_{r}$, i.e.,
$i+1$. This is in perfect agreement with the three marked degeneracy
points shown in Fig. 1(a).  It is now also possible to explain why
the level crossings in the middle of the spectrum in Fig. 1(a) can
involve three states. This is because: (i) if $N$ is even and
$g_0=0$, $H_{\text{eff}}$ always has a zero eigenvalue due to its
tri-diagoal structure; and (ii) when the dimension of $h_{r}$ and
$h_{l}$ is odd, they can present two additional zero eigenvalues.
Interestingly, for sufficiently large $g_1$, such agreement between
theory and numerics may persist for the intermediate-frequency case
in Fig. 1(b). For example, the degeneracies marked by the three
vertical lines in Fig. 1(b) occur at $g_1/\omega\sim 2.007$,
$2.133$, and $2.986$. These values, when multiplied by $(N-(2i+1))$
for $i=0,1,2$, respectively, are the 6-th or 5-th root of ${\cal
J}_0(x)$. Note however, that the subtle crossing behavior depicted in the
inset of Fig. 1(b) is beyond our theory.

With a normalized population imbalance $\langle S \rangle \equiv\
2\langle J_z\rangle/N$,  Fig. \ref{Fig:population}(a) shows the
numerically exact population dynamics associated with the three
marked points in Fig. 1(a). The initial state is that all particles
are in the left well. In the first case for $g_{1}/\omega\sim
0.267$, $\langle S\rangle$ is seen to stay at almost unity and hence
in essence the tunneling between the two wells is {completely}
suppressed. In the second case for $g_{1}/\omega\sim 0.344$, our
theory predicts that the tunneling suppression occurs only when
$\langle S \rangle$ becomes 0.8. As seen in Fig. 2(a), $\langle
S\rangle$ indeed oscillates between $0.8$ and 1.0. Similarly, in the
third case, $\langle S\rangle$ oscillates between 1.0 and $\sim
0.6$, confirming our theory that the tunneling stops if two
particles are already released to the right well. Excellent
agreement is obtained at other level degeneracy points. These
features signify one key aspect of our many-body CDT: it depends
sensitively on the number of particles that have already tunneled.
With the same initial condition, in Fig. 2(b) we also show the three
intermediate-frequency cases marked earlier by the vertical lines in
Fig. 1(b). The associated population dynamics still agrees with our
theory. Because the predicted CDT points are independent of the
actual tunneling rate $v$, we found that even if an oscillation in
$v$ is considered (which can be induced by the modulation in
$g(t)$), analogous results can be obtained. For the case shown in
the inset of Fig. 1(b), which is beyond our high-frequency theory,
almost complete population delocalization is observed in the inset
of Fig. 2(b).

\begin{figure}
\begin{center}
\begin{tabular}{lc}
\includegraphics[width=7.5cm,height=6.cm]{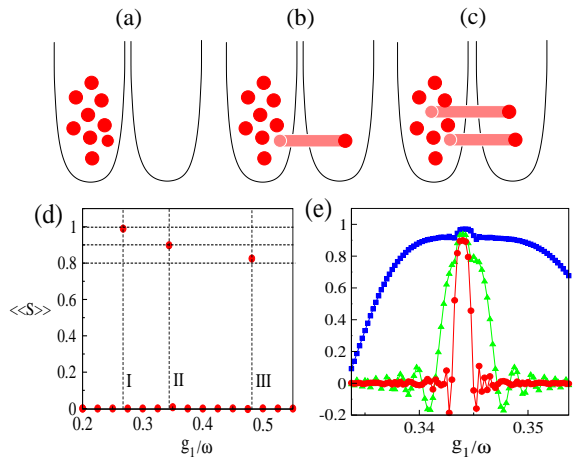}
\end{tabular}
\caption{(Color online) (a)-(c): Schematic picture of particle number-dependent
CDT. In case (a) no particle tunnels, and in cases (b) and (c) only
one or two particles tunnels and then CDT occurs. (d): Long-time
average of $\langle \langle S \rangle \rangle$ vs $g_{1}/\omega$ for
$\omega=40$ and $g_0=0$, with $g_1/\omega$ scanned at a rather low
resolution. (e): Same as in (d), but with $g_1/\omega$ scanned in
small steps around the regime of 0.343, with $g_{0} /\omega= 1/144$
(squares), $g_{0} /\omega= 1/288$ (triangles), and $g_{0}=0$
(circles).  The total time used for averaging is $20000$ in
dimensionless units. } \label{Fig:quantization}
\end{center}
\end{figure}

Figure 3(a)-(c) schematically illustrates the three representative
CDT cases studied in Fig. 1(a) and Fig. 2(a). In Fig. 3(a), no
particle is allowed to tunnel; and in Fig. 3(b) and 3(c) one or two
particles have tunneled and then CDT occurs. Consistent with this
picture, Fig. 3(d) depicts the numerically time-averaged $\langle
S\rangle$, denoted $\langle \langle S\rangle\rangle$, as a function
of $g_1/\omega$, for $N=10$ and $\omega=40$.  It is seen that as
$g_1/\omega$ is scanned at a rather low resolution, the value of
$\langle \langle S\rangle\rangle$ is either zero or close to
``magic" nonzero numbers ($\sim 1.0$, $\sim 0.9$, and $\sim 0.8$,
$\cdots$). Figure 3(a)-3(c) also provoke us to re-interpret our
theoretical finding. In particular, for $g_0=0$, the energy
difference between the two configurations in Fig. 3(a) and Fig. 3(b)
(Fig. 3(b) and Fig. 3(c)) is $g_1\hbar\cos(\omega t)[N-(2i+1)]$ with
$i=0$ ($i=1$).  As such, even though there is no direct modulation
of the energy bias between the two wells, the oscillation in $g(t)$
still causes a modulation of the {\it effective} bias between
different configurations. With this interpretation we are able to
re-derive Eq. (\ref{con2}) by analogy to  standard single-particle
CDT theory. This analogy also makes clear that the precise condition
for our high-frequency approximation should be $\tilde{v} \ll
\max{(\omega,\sqrt{\epsilon \omega})}$ \cite{Hanggirev}, where
$\tilde{v}\equiv v \sqrt{(N-i)(i+1)}$ is the coupling strength
between states $|N/2-i\rangle$ and $|N/2-i-1\rangle$, and
$\epsilon\equiv |g_1[N-(2i+1)]|$ is the amplitude of the effective
bias. For $i\ll N$, this condition becomes $v\ll
\max{(\frac{\omega}{\sqrt{N}}, \sqrt{g_1 \omega})}$. This well
explains our early observation from Fig. 1(b) that as $g_1$
increases, the spectral pattern in intermediate-frequency cases
starts to resemble those in Fig. 1(a) and becomes more perspicuous
with our theory.

Let us now turn to cases with nonzero $g_0$. If the system is close
to a CDT point, ${\cal J}_{0}\left[g_1(2J_z+1)/\omega\right]$ is
small, the $g_0$ term in $H_{\text{eff}}$ will dominate and hence
the associated self-trapping effect may induce a strong population
imbalance on a very long time scale. Taking one small window of
$g_1/\omega$ in Fig. 3(d) as an example, we compare $g_0\ne 0$ with
$g_0=0$ cases in Fig. 3(e). Clearly, as $g_0$ increases, the width
of the $\langle\langle S\rangle\rangle$ profile in Fig. 3(e)
increases significantly. This interplay between self-trapping and
CDT is analogous to that in a two-mode optical waveguide system
where the mode bias is periodically modulated \cite{wubiao}. The
peak value of $\langle\langle S\rangle\rangle$ is also seen to
change with $g_0$. We conclude that on one hand a small nonzero
$g_0$ is beneficial to experiments because it reduces the
sensitivity of $\langle\langle S\rangle\rangle$ to the exact values
of $g_1/\omega$; on the other hand, {however}, the predicted
particle-number dependent CDT effect may be buried by self-trapping
if $g_0$ is too large.

Our findings have potential applications in probing and exploring
genuine quantum coherence in BEC. In particular, the dynamics under
a pre-established CDT condition may be dramatically changed upon
adding bosons to the system.  As an example we consider the CDT
point I  in Fig. 1(a), where ${\cal
J}_{0}\left[g_1(N-1)/\omega\right]=0$
 for $N=10$. If we let
$N\rightarrow N+2$ by adding two particles to the left well, then
because we still have ${\cal
J}_{0}\left[g_1((N+2)-3)/\omega\right]=0$, Eq. (\ref{con2}) suggests
that the CDT will be re-established after one particle is tunneled
to the right well. However, if we let $N\rightarrow N+1$ by adding
only one particle to the left well, then because $(N-1)$ cannot be
written as $(N+1)-(2i+1)$ for any $i$, ${\cal
J}_{0}\left[g_1((N+1)-(2i+1))/\omega\right]$ is in general nonzero
for fixed $g_1/\omega$, and as a result all the particles start to
tunnel back and forth between the two wells. The minor difference
between adding an even and adding an odd number of bosons is thus
greatly amplified by CDT, a prediction also confirmed by our
numerical experiments. Similar behavior is obtained if more
particles are added to the system.  This odd-even sensitivity to the
particle number is absent in any mean-field theory of a BEC,
providing a possible means for accurate counting or efficient
filtering of the number of bosons. Certainly, as implied by the
results in Fig. 3(e), this is possible only if $g_0$ is sufficiently
small such that the $\langle\langle S\rangle\rangle$ profiles
associated with different particle numbers do not overlap.

This work is supported by WBS grant Nos. R-144-050-193-101/133 and
R-144-000-195-101 (J.G.), and by German Excellence Initiative via
the \textit {Nanosystems Initiative Munich} (NIM) (P.H.).


%


\begin{thebibliography}{99}

\bibitem{CDT1991}
F. Grossmann {\it et al}.,  
Phys. Rev. Lett. {\bf 67}, 516 (1991); F. Grossmann and P. H\"anggi,
Europhys. Lett. {\bf 18},  571 (1992).

\bibitem{Hanggirev} M. Grifoni  and P. H\"{a}nggi, Phys. Rep. {\bf
304}, 229 (1998).



\bibitem{devalle}
G. Della Valle {\it et al.}, \prl{\bf 98}, 263601 (2007); H. Lignier
{\it et al}., \prl{\bf 99}, 220403 (2007).

\bibitem{oberthaler}
E. Kierig {\it et al}., \prl{\bf 100},  190405 (2008).

\bibitem{creffield} 
C.E. Creffield and T.S. Monteiro, \prl{\bf 96},
210403 (2006); C.E. Creffield, \pra{\bf 75}, 031607 (2007).

\bibitem{holthaus} A. Eckardt and M. Holthaus, \prl{\bf 101}, 245302
(2008).

\bibitem{Erimondo}A. Zenesini {\it et al.}, \prl{\bf 102}, 100403
(2009); A. Eckardt {\it et al.}, \pra{\bf 79}, 013611 (2009).

\bibitem{kevre} P.G. Kevrekidis {\it et al.}, 
\prl{\bf 90}, 230401 (2003); Kh. Abdullaev {\it et al.}, 
\pra{\bf 68}, 053606 (2003); H. Saito and M. Ueda,
\prl{\bf 90}, 040403 (2003).

\bibitem{qi} Q. Zhang, P. H\"{a}nggi, and J.B. Gong, \pra{\bf 77}, 053607
(2008); and references therein.
\bibitem{naturepapers}S. F\"{o}lling, {\it et al.}, Nature {\bf 448}, 1029 (2007);
M. Anderlini, {\it et al.}, Nature {\bf 448}, 452 (2007).
\bibitem{korsch}This is already exploited in
a delta-kicked system; see M.P. Strzys, E.M. Graefe, and H.J.
Korsch, New. J. Phys. {\bf 10}, 013024 (2008).

\bibitem{kohler}S. Kohler, J. Lehmann, and P. H\"{a}nggi, Phys. Rep. {\bf 406},
379 (2005); see pp. 401-402.
\bibitem{wubiao}X. Luo, Q. Xie, and B. Wu, \pra{\bf 76}, 051802
(2007).

\end{thebibliography}
\end{document}